\documentclass[3p,times,twocolumn]{elsarticle}

\usepackage{ecrc}
\usepackage{multirow}

\volume{00}

\firstpage{1}

\journalname{Nuclear Physics B Proceedings Supplement}

\runauth{Raymond R. Volkas}

\jid{nuphbp}

\jnltitlelogo{Nuclear Physics B Proceedings Supplement}

\usepackage{amssymb}

\usepackage[figuresright]{rotating}

\begin{document}

\begin{frontmatter}



\dochead{}

\title{Unified origin for visible and dark matter in a baryon-symmetric universe from a first-order phase transition}


\author{Raymond R. Volkas}
\ead{raymondv@unimelb.edu.au}

\address{ARC Centre of Excellence for Particle Physics at the Terascale, School of Physics, The University of Melbourne, Victoria 3010, Australia}

\begin{abstract}
In a baryon-symmetric universe, the baryon asymmetry observed for visible matter is matched by an equal and opposite asymmetry for dark matter, thereby closely connecting the number densities of both types of matter.  This is a necessary step towards the goal of explaining the mystery of why the visible and dark matter densities are observed to be similar.  In this talk, a way of producing such a universe from bubble nucleation during a first-order phase transition is reviewed.  The process is an analog of electroweak baryogenesis.
\end{abstract}


\end{frontmatter}


\section{Introduction}
\label{sec:intro}

The fact that the cosmological mass density of ordinary or visible matter (VM) today is only about a factor of five different from the mass density of dark matter (DM),
\begin{equation}
\Omega_d \simeq 5 \Omega_v,
\end{equation}
suggests a common origin for both.  In this talk, I review work performed with K. Petraki and M. Trodden~\cite{PTV} on a model that (partially) explains this fact from the unified production of VM and DM through the agency of bubble nucleation dynamics during a first-order phase transition in the early universe, an analog of the electroweak baryogenesis mechanism~\cite{EWB}.  It is a particular scenario drawn from the special class of asymmetric DM models~\cite{ADM} which feature a ``baryon-symmetric universe''~\cite{DW}.  This idea is explained below.

We start by discussing the popular Weakly Interacting Massive Particle (WIMP) hypothesis for DM.  The idea is that the DM is a new particle with an electroweak-scale mass, typically a Majorana fermion such as the neutralino of supersymmetric theories, whose cosmological density is determined by a weak-scale annihilation cross-section.  As the temperature of the universe drops below the WIMP mass, the creation of WIMPs becomes energetically disfavored and their density drops through Boltzmann suppression until they decouple from the VM plasma.  With a weak-scale annihilation cross-section, it turns out that the WIMPs decouple when their mass density is in the correct range to explain the cosmological DM observations.  This scenario is considered to be attractive because the existence of weakly-interacting electroweak-scale particles is motivated by independent particle physics problems.  The neutralino, for example, is required by the supersymmetric solution of the gauge hierarchy problem.

However, the WIMP hypothesis leaves unexplained why the DM density is so similar to the VM density.  The latter is \emph{not} determined by the physics of weak-scale annihilation.  Rather, it is due to the existence of the cosmological baryon asymmetry: some as-yet unknown physics in the early universe caused the baryons to outnumber the antibaryons by about one part in $10^{10}$.  The baryon-antibaryon annihilations which deplete those species' number densities at temperatures below their masses switch off not because the interactions become too weak, but because the baryons eventually have no antibaryons to annihilate with.  The excess baryons form the VM in the universe today.  Weak-scale annihilation that becomes ineffective through decoupling compared with strong-sector annihilation that switches off because of a particle-antiparticle asymmetry: these are very different processes, involving different forces and circumstances, yet in the WIMP paradigm they are required to coincidentally produce similar relic mass densities.  Acknowledging this necessary coincidence motivates that an alternative to the WIMP hypothesis be given serious consideration.

The obvious alternative is simply to suppose that the DM density is also due to a particle-antiparticle asymmetry, and that there is a dynamical connection between the visible and dark sectors that makes the asymmetries similar, or possibly even identical.  This is the ``asymmetric DM'' hypothesis.  The idea is that the DM consists of stable relics from a hidden sector that possesses an analog of baryon number.  The dark sector may be its own complicated world, described by some kind of gauge theory.  For the special case of ``mirror matter''~\cite{mirror}, the dark gauge theory is isomorphic to the standard model (SM), but in general it is different.  In the low-energy world of the late universe in which we live, we came to know of the other sector through the necessarily common gravitational interaction.  But the similar densities suggest that there are also non-gravitational connections to be discovered between our world and the dark world.  In asymmetric DM models, the similar VM and DM asymmetries point to a DM particle mass in the few GeV regime.  The DAMA, CoGeNT and CRESST anomalies provide some encouragement to take this seriously~\cite{DAMA}.

A baryon-symmetric universe is one that contains a DM asymmetry that is exactly the opposite of the VM asymmetry, so it is a special case of asymmetric DM.  It is attractive because the strong connection between the two asymmetries is driven by a symmetry principle rather than by the specifics of some dynamical scheme.  Indeed, many different theories can produce a baryon-symmetric universe, and this talk is about only one of them.  Establishing a connection between the visible and dark asymmetries is a necessary step towards the goal of a complete understanding of the similar visible and dark \emph{mass} densities.  One also needs a theory for the origin of the few-GeV DM mass scale, but that important problem is beyond the scope of this talk.

Here is the symmetry principle~\cite{pangenesis}:  Call the ordinary baryon number $B_1$, and let $B_2$ standard for the DM analog.  Form the orthogonal linear combinations
\begin{eqnarray}
B &\equiv& B_1 - B_2 \nonumber\\
X &\equiv& B_1 + B_2\ .
\end{eqnarray}
Now demand that $B$ is either an exactly or essentially conserved quantum number, while $X$ is violated at high energies and in the early universe. (By ``essentially conserved'' we mean we remain open to the existence of extremely weak non-conserving processes, such as very slow proton decay, but these are so weak that they are cosmologically irrelevant.)  Dynamics that violates $X$ through out-of-equilibrium processes that also violate $C$ and $CP$ will, according to the Sakharov analysis~\cite{Sakharov}, produce an $X$ asymmetry.  But no $B$ asymmetry develops, so we have that
\begin{equation}
\Delta B_1 = \Delta B_2 = \Delta X/2\ .
\end{equation}
The quantity $B$ may be thought of as a generalized baryon number.  At low energies and in the late universe, $X$ violation becomes very weak, and $B_1$ and $B_2$ become individually conserved.  In the visible sector, $B_1$ conservation ensures the (essential) stability of protons and this accounts for the bulk of the VM density.  Electric charge and angular momentum conservation also make the electron and, respectively, the lightest neutrino stable.  In the dark sector, $B_2$ conservation ensures the stability of at least one species.  For other reasons, there may be other stable dark species.  Non-perturbative sphaleron effects may reprocess some of these asymmetries in one or both sectors.

The requirement that $B$ is never violated is a strong one.  To justify such an imposition, it is natural to take U(1)$_B$ to be a gauge symmetry, while U(1)$_X$ is global.  In that case, anomaly-freedom suggests that $B$ should rather be replaced by $B-L$, which is indeed what we do below.  Any such model must be constructed to ensure that a global U(1)$_B$ is produced after the gauged U(1)$_{B-L}$ is spontaneously broken.  There is a standard way to do this, as explained in Ref.~\cite{PTV}.

The first consideration in building a model for a baryon-symmetric universe is the dynamics of $X$ asymmetry generation.  We may borrow the essential idea from each known way of generating an ordinary baryon asymmetry, and reuse it to produce a nonzero $X$.  Well-studied schemes include out-of-equilibrium heavy particle decays, Affleck-Dine dynamics~\cite{AD}, and bubble nucleation during a first-order phase transition (as used in electroweak baryogenesis).  We employ the latter here.  See Refs.~\cite{pangenesis, Bsym} for papers on baryon-symmetric universe models.

We first briefly recall how electroweak baryogenesis works.  Above the critical temperature for the electroweak phase transition, the Higgs vacuum expectation value is zero and weak sphalerons induce certain rapid $B_1$-violating processes.  In electroweak baryogenesis, the phase transition is required to be strongly first order:  bubbles of broken phase are nucleated at the critical temperature within the pre-existing symmetric phase universe.  Sufficiently large bubbles grow rapidly, and the passage of the bubble walls produces the required departure from thermal equilibrium.  Top quark interactions with the moving walls violate $C$ and $CP$, and sphaleron-induced $B_1$ violation continues in the regions of symmetric phase.  These ingredients work together to produce a $B_1$-asymmetric universe once the phase transition has been completed and the universe is fully in the broken phase.  In the next section we describe how these dynamics may be recast to produce an $X$-asymmetric, but $B$-symmetric, universe.

\section{The model}
\label{sec:model}

The model consists of three sectors: visible, ``generative'' and dark.  The visible sector is simply the SM, extended to include a gauged U(1)$_{B-L}$, which means we include three right-handed neutrinos in the visible particle spectrum.  It also includes a vector-like sterile fermion $f$ whose role is explained below.  The generative sector, to be described shortly, is the engine room of $X$ asymmetry creation.  Once nonzero $X$ has been created by the dynamics of that sector, it is transferred to the visible sector and the dark sector.  The dark sector of our model has the gauge symmetry U(1)$_D$, as well as the U(1)$_{B-L}$ it shares with the visible sector.  The local U(1)$_D$ symmetry is not spontaneously broken, so it can be thought of as a dark sector version of electromagnetism.  It is needed to annihilate away the symmetric part of the dark plasma, leaving the excess dark ``antibaryons" as the relic DM. The symmetries and field content of our model are summarized in Table~\ref{tab:charges}.

\begin{table}[t]
\begin{center}
\begin{tabular}{|c|l|c|c|c|c|}
\hline 
\multirow{2}{*}{} & \multirow{2}{*}{Fields} & SU(2)$_G$ & U(1)$_{B-L}$ & U(1)$_X$  & U(1)$_D$  \\
                         &                            & gauged    & gauged       & anom. & gauged   \\  
\hline \hline
                         & $\psi_L$                    &  2        &  0           & -2        & 0  \\
g               & $\psi_{1,2R}$       &  1        &  0           & -2        & 0   \\ 
                         & $\varphi$                      &  2        &  0           &  0        & 0   \\ \hline
\multirow{2}{*}{v} & $ f_{L,R}$                 &  1        & -1           & -1        & 0   \\ 
                         & $ \nu_R$                    &  1        & -1           & -1        & 0  \\ \hline
                         & $\chi$                       &  2        &  1           & -1        & 0   \\ 
d                     & $\xi_{L,R}$                &  2        &  0           &  0        & 1   \\ 
                         & $\zeta_{L,R}$              &  1        & -1           &  1        & 1   \\ 
\hline
\end{tabular}
\end{center}
\caption{The charge assignments under the new symmetries in the model, where g, v, and d denote the generative, visible and dark sectors, respectively. The model consists of the SM fields, the fields in this table and another scalar that breaks the gauged U(1)$_{B-L}$ into a global baryon number symmetry. Three right-handed neutrinos are needed to cancel the cubic $B-L$ anomaly, and an even number of families of the $\psi$ fermions is required to ensure the absence of a Witten anomaly.
The generative-sector field $\varphi$ gains a nonzero VEV during a first-order phase transition, while the dark-sector scalar $\chi$ always has a zero VEV.  The fields $f$ and $\chi$ facilitate the transfer of the $X$ asymmetry to the visible and dark sectors.  The stable DM is formed from the vector-like fermions $\xi$ and $\zeta$, which form ``atomic'' two-body bound states through the U(1)$_D$ dark electromagnetism.
}
\label{tab:charges}
\end{table}

The generative sector is constructed to make U(1)$_X$ anomalous, with sphaleron transitions associated with a new SU(2)$_G$ gauge interaction providing the necessary violation of $X$ conservation.  This structure is borrowed from electroweak baryogenesis, where the U(1)$_{B_1}$ current has a weak-isospin SU(2)$_L$ anomaly, and weak sphalerons mediate processes that violate $B_1$ conservation.  The fermions $\psi$ in the generative sector have a similar chirality structure to that of ordinary quarks and leptons with respect to weak interactions:  the left-handed $\psi$'s are placed in an SU(2)$_G$ doublet, while the two right-handed components are singlets.  Both chiral components carry $X$ charges, but the fact that only the left-handed ones feel the SU(2)$_G$ interactions causes U(1)$_X$ to be anomalous.  There must be an even number of families of the $\psi$ fermions to ensure absence of a Witten anomaly~\cite{Witten}, the minimum number being two.  The generative sector also contains an SU(2) scalar doublet $\varphi$, whose role is to spontaneously break SU(2)$_G$ in a strongly first-order phase transition.  Because the SU(2)$_G$ is an exotic gauge interaction, we can make the phase transition as strongly first-order as we wish by choosing the scalar potential parameters appropriately.  Recall that this freedom is not available for SU(2)$_L$ breaking in the SM, because the parameter choice required to produce a standard Higgs boson of the observed mass is not consistent with a first-order electroweak phase transition (so electroweak baryogenesis can work only within some SM extensions).  The two families of generative $\psi$ fermions Yukawa couple to $\varphi$ in a $C$- and $CP$-violating way.

We now describe the dynamics of asymmetry creation and transfer to the visible and dark sectors, which serves also to explain the roles of the exotic particles $f$, $\chi$, $\xi$ and $\zeta$.  At a critical temperature somewhat above the electroweak scale, the $\varphi$ field develops a nonzero vacuum expectation value (VEV) through a strongly first-order phase transition, which proceeds via bubble nucleation of the broken phase.  In the regions of unbroken phase SU(2)$_G$ sphaleron effects cause rapid $X$-violating processes to occur, while such effects are absent in the regions of broken phase.  The $\psi$ fermions interact with the fast-moving $\varphi$ bubble walls in a $C$- and $CP$-violating way.  Emulating electroweak baryogenesis, after the phase transition has been completed these circumstances leave a plasma with an $X$ asymmetry, carried by the $\psi_L$ particles.  The Yukawa interactions $\bar{\psi}_L \chi f_R + {\rm H.c.}$ then transfer the asymmetry via $\psi_L$ decay to the visible sector vector-like sterile fermions $f$ and the dark sector scalar $\chi$.  Gauge invariance dictates the multiplet assignments of these particles.  Since $f$ and $\chi$ carry equal and opposite $B-L$ charges, the baryon-symmetric feature is preserved through the transfer.  Note that $\chi$ does not carry dark electromagnetic charge.

The fields $f_R$ Yukawa couple to the SM lepton doublets through the standard Higgs doublet, reprocessing the asymmetry it carries into those fields.  Through other SM interactions, and weak sphalerons, the asymmetry spreads throughout the visible sector.

The dark sector must be larger than the scalar multiplet $\chi$, because that field is necessarily immune from the dark electromagnetic force we need to use to annihilate away the symmetric part.  This requires the existence of the U(1)$_D$-charged vector-like fermions $\xi$ and $\zeta$, which Yukawa couple to $\chi$ through the interactions 
\begin{equation}
\bar{\xi}_L \chi \zeta_R + (L \leftrightarrow R) + {\rm H.c.}
\end{equation}
We choose the free mass parameters to make $\chi$ decouple and decay through this interaction into the $\xi$ and $\zeta$ dark fermions.  The unbroken dark electric charge conservation works with the now essentially restored $B_2$ conservation to make all of these fermions stable (assuming there is only one family).  The symmetric part annihilates into dark photons, leaving an excess of $\bar{\xi}$ and $\zeta$ particles to be the relic DM.  In fact, the DM today is in the form of ``atomic'' bound states of $\bar{\xi}$ and $\zeta$, with the binding due to the U(1)$_D$ force.  The DM is thus a kind of ``dark hydrogen'', though there are some dissimilarities: there is no dark nuclear physics, and there need not be such a large difference in mass between $\bar{\xi}$ and $\zeta$ as there is between the proton and the electron.  To obtain the observed DM mass density, we must have
\begin{equation}
m_\xi + m_\zeta \simeq 0.3 m_p \frac{\Omega_d}{\Omega_v} \simeq 1.5\ {\rm GeV},
\end{equation}
where $m_p$ is the proton mass and the $0.3$ factor is due to weak sphaleron reprocessing in the visible sector.

\section{Constraints}
\label{sec:constraints}

The parameters in the model must be chosen to obey certain cosmological and astrophysical constraints.

One important requirement is that the dark photonic radiation created from the annihilation of the symmetric part of the dark plasma does not spoil big bang nucleosynthesis through an unacceptably large increase in the expansion rate of the universe over the standard scenario.  The amount of dark radiation is governed by the number of degrees of freedom in the dark sector that undergo the annihilations.  The constraint is therefore an upper bound~\cite{PTV}
\begin{equation}
g_{d, {\rm dec}} \lesssim 18.6 \left( \frac{g_{v,{\rm dec}}}{110.25} \right)
\end{equation}
where $g_{d, {\rm dec}}$ and $g_{v, {\rm dec}}$ are the degree of freedom counts in the dark and visible sectors, respectively, at the time the two sectors decouple from each other in the early universe.  The 18.6 figure comes from bounding the extra radiation to be at most the equivalent of one additional neutrino flavor.  The dark sector of the model described above has $g_{d, {\rm dec}} = 16.5$, so it passes the test.

Another constraint arises from ensuring successful large scale structure formation.  As well as explaining various gravitational anomalies in the present day universe, such as flat galactic rotation curves, DM is also needed to seed the growth of structure in the early universe.  In the standard cold DM scenario, structure begins to form in the DM soup as soon as matter starts to dominate over radiation.  The visible matter is still ionized at this stage, and the relatively strong electromagnetic force prevents the gravitational growth of structure in that sector.  Baryonic structure growth is therefore delayed until neutral atoms form and the ordinary photons decouple, but by then there are already significant DM overdensities to facilitate the growth of inhomogeneities in ordinary matter.  To ensure that this successful scenario is not spoiled, we must require that dark photon decoupling happens no later than the point of matter-radiation equality.  It turns out that this leads to constraints on the reduced $\bar{\xi}, \zeta$ mass and the dark fine structure constant $\alpha_D$ that are quite easy to satisfy~\cite{PTV}.

Finally, the Bullet cluster observations are argued to provide an upper bound on the DM self-interaction cross-section $\sigma$~\cite{Bullet}:
\begin{equation}
\frac{\sigma}{m_d} \lesssim 1\ {\rm cm}^2/{\rm g}\ ,
\end{equation}
where $m_d$ is the mass of the DM particle.  For the atomic DM of the described model, the cross-section is determined by the geometric size of the DM atoms.  To ensure sufficiently tightly bound DM atoms, the bound
\begin{equation}
\alpha_D > 0.3
\end{equation}
is indicated~\cite{PTV}.  On the face of it, this is the strongest constraint on the model.

\section{Signatures}
\label{sec:signatures}

The model possesses features that can be searched for in collider and direct DM detection experiments.  On the collider side, the gauged U(1)$_{B-L}$ implies the existence of a $Z'$ boson that has a significant invisible width into DM.  The $B-L$ breaking scale could be low enough for such a particle to be observable at the LHC. The scalar sector of the theory is quite rich, and will in general lead to modifications of SM Higgs boson properties and of course the existence of additional spin-0 particles.  Direct DM detection experiments are relevant for this model because of the $Z'$ interaction.  One may compute that the $Z'$-induced spin-independent DM-nucleon cross-section is~\cite{PTV}
\begin{equation}
\sigma^{\rm SI} \simeq 10^{-44}\ {\rm cm}^2\ \left( \frac{g_{B-L}}{0.1} \right)^4 \left( \frac{0.7\ {\rm TeV}}{M_{B-L}} \right)^4\ ,
\end{equation}
where $g_{B-L}$ and $M_{B-L}$ are the $Z'$ gauge coupling constant and mass, respectively.  For reasonable choices of these parameters, the bounds from XENON100~\cite{XENON100} and CDMS II~\cite{CDMSII} are easily met.  The DM-nucleon cross-section may be enhanced in a variant of this model, where U(1)$_D$ is spontaneously broken and the now massive dark photon kinetically mixes with the ordinary photon.

\section{Conclusion}
\label{sec:conclusion}

The similar visible and dark matter densities suggest a common origin for both types of matter, as can be obtained in asymmetric DM models in general and the baryon-symmetric special cases in particular.  A baryon-symmetric universe has the DM carrying a generalized baryon asymmetry that cancels the asymmetry in the visible sector, and leads to a very tight relationship between ordinary and DM enforced by a simple symmetry principle.  In this talk, a baryon symmetric universe was described that has the equal and opposite visible and dark asymmetries produced by an analog of electroweak baryogenesis.  Cosmological and astrophysical constraints were examined, and signatures for collider and direct DM detection experiments very briefly discussed.

\section*{Acknowledgments}
This work was supported in part by the Australian Research Council.  The author thanks his excellent collaborators K. Petraki and M. Trodden, and warmly acknowledges the hospitality and support of the organizers of the CosPA 2012 symposium.


\begin{thebibliography}{00}

\bibitem{PTV} K. Petraki, M. Trodden and R. R. Volkas, JCAP 1202 (2012) 044.

\bibitem{EWB} For a review see, A. Riotto and M. Trodden, Ann.\ Rev.\ Nucl.\ Part.\ Sci.\ 49 (1999) 35.

\bibitem{ADM} For a recent review see, H. Davoudiasl and R. N. Mohapatra, New J. Phys.\ 14 (2012) 095011, and see~\cite{PTV} for an extensive list of references to asymmetric dark matter models.

\bibitem{DW} S. Dodelson and L. M. Widrow, Phys.\ Rev.\ Lett.\ 64 (1990) 340; Phys.\ Rev.\ D 42 (1990) 326; Mod.\ Phys.\ Lett.\ A5 (1990) 1623.

\bibitem{mirror} See, for example, R. Foot, H. Lew and R. R. Volkas, Phys.\ Lett.\  B 272 (1991) 67; Z. Berezhiani, D. Comelli, F. L. Villante, Phys.\ Lett.\  B 503 (2001) 362; A. Yu.\ Ignatiev and R. R. Volkas, Phys.\ Rev.\ D 68 (2003) 023518; R. Foot and R. R. Volkas, Phys.\ Rev.\ D 68 (2003) 021304; Phys.\ Rev.\ D 69 (2004) 123510; Z. Berezhiani et al., Int.\ J. Mod.\ Phys.\  D 14 (2005) 107; P. Ciarcelluti, Int.\ J. Mod.\ Phys.\ D 14 (2005) 187; Int.\ J. Mod.\ Phys.\ D 14 (2005) 223; R. Foot, Phys.\ Rev.\ D 82 (2010) 095001; Phys.\ Rev.\ D 86 (2012) 023524; arXiv:1211.1500; J.-W. Cui et al., Phys.\ Rev.\ D 85 (2012) 096003.

\bibitem{DAMA} For citations to the original literature and an analysis of how these results may be reconciled with hidden sector dark matter, see R. Foot, arXiv:1209.5602.

\bibitem{pangenesis} N. F. Bell, K. Petraki, I. M. Shoemaker and R. R. Volkas. Phys.\ Rev.\  D 84 (2011) 123505;  L. J. Hall, J. March-Russell and S. M. West, arXiv:1010.0245.

\bibitem{Sakharov} A. D. Sakharov, Pisma, Zh.\ Eksp.\ Theor.\ Fiz.\ 5 (1967) 32.

\bibitem{AD} I. Affleck and M. Dine, Nucl.\ Phys.\ B 249 (1985) 361; M. Dine, L. Randall and S. Thomas, Nucl.\ Phys.\ B 458 (1996) 291.

\bibitem{Bsym} V. A. Kuzmin, Phys.\ Part.\ Nucl.\ 29 (1998) 257;
R. Kitano and I. Low, Phys.\ Rev.\ D71 (2005) 023510; hep-ph/0503112;
P.-H. Gu, Phys.\ Lett.\ B657 (2007) 103;
P.-H. Gu, U. Sarkar and X. Zhang, Phys.\ Rev.\ D80 (2009) 076003;
H. An, S.-L. Chen, R. N. Mohapatra and Y. Zhang, JHEP 03 (2010) 124;
H. Davoudiasl, D. E. Morrissey, K. Sigurdson and S. Tulin,  Phys.\ Rev.\ Lett.\ 105 (2010) 211304;
P.-H. Gu, M. Lindner, U. Sarkar and X. Zhang, Phys.\ Rev.\ D83 (2011) 055008;
J. J. Heckman and S.-J. Rey, JHEP 1106 (2011) 120;
D. H. Oaknin and A. Zhitnitsky, Phys.\ Rev.\ D71 (2005) 023519;
G. R. Farrar and G. Zaharijas, Phys.\ Rev.\ Lett.\ 96 (2006) 041302;
C. Cheung and K. M. Zurek, Phys.\ Rev.\ D84 (2011) 035007;
B. von Harling, K. Petraki and R. R. Volkas, JCAP 1205 (2012) 021;
J. March-Russell and M. McCullough, arXiv:1106.4319;
K. Kamada and M. Yamaguchi, arXiv:1201.2636.

\bibitem{Witten} E. Witten, Phys.\ Lett.\ B 117 (1982) 324.

\bibitem{Bullet}
M. Markevitch et. al., Ap.\ J. 606 (2004) 819;
S. W. Randall et al, Ap.\ J. 679 (2008) 1173.

\bibitem{XENON100}
XENON100 Collaboration, E. Aprile et. al., Phys.\ Rev.\ Lett.\ 107 (2011) 131302.

\bibitem{CDMSII}
CDMS-II Collaboration, Z. Ahmed et. al., Phys.\ Rev.\ Lett.\ 106 (2011) 131302.












\end{thebibliography}
\end{document}